\documentclass[format=sigconf,nonacm]{acmart}
\usepackage[ruled,linesnumbered]{algorithm2e}
\usepackage{bm}
\usepackage{makecell}
\usepackage{multirow}
\renewcommand\footnotetextcopyrightpermission[1]{}

\AtBeginDocument{%
  \providecommand\BibTeX{{%
    \normalfont B\kern-0.5em{\scshape i\kern-0.25em b}\kern-0.8em\TeX}}}

\begin{document}

\title{Scientific and Technological Information Oriented Semantics-adversarial and Media-adversarial Cross-media Retrieval}

\author{Ang Li}
\affiliation{
	\institution{Beijing University of Posts and Telecommunications, Beijing Key Laboratory of Intelligent Telecommunication Software and Multimedia}
	\city{Beijing}
	\country{China}
	\postcode{100876}
}
\email{david.lee@bupt.edu.cn}

\author{Junping Du}
\authornote{Corresponding author}
\affiliation{
	\institution{Beijing University of Posts and Telecommunications, Beijing Key Laboratory of Intelligent Telecommunication Software and Multimedia}
	\city{Beijing}
	\country{China}
	\postcode{100876}
}
\email{junpingdu@126.com}

\author{Feifei Kou}
\affiliation{
	\institution{Beijing University of Posts and Telecommunications, Beijing Key Laboratory of Intelligent Telecommunication Software and Multimedia}
	\city{Beijing}
	\country{China}
	\postcode{100876}
}
\email{koufeifei000@126.com}

\author{Zhe Xue}
\affiliation{
	\institution{Beijing University of Posts and Telecommunications, Beijing Key Laboratory of Intelligent Telecommunication Software and Multimedia}
	\city{Beijing}
	\country{China}
	\postcode{100876}
}
\email{xuezhe@bupt.edu.cn}

\author{Xin Xu}
\affiliation{
	\institution{Beijing University of Posts and Telecommunications, Beijing Key Laboratory of Intelligent Telecommunication Software and Multimedia}
	\city{Beijing}
	\country{China}
	\postcode{100876}
}
\email{xinxu@bupt.edu.cn}

\author{Mingying Xu}
\affiliation{
	\institution{Beijing University of Posts and Telecommunications, Beijing Key Laboratory of Intelligent Telecommunication Software and Multimedia}
	\city{Beijing}
	\country{China}
	\postcode{100876}
}
\email{xumingying0612@126.com}

\author{Yang Jiang}
\affiliation{
	\institution{Beijing University of Posts and Telecommunications, Beijing Key Laboratory of Intelligent Telecommunication Software and Multimedia}
	\city{Beijing}
	\country{China}
	\postcode{100876}
}
\email{jugyang@163.com}

\renewcommand{\shortauthors}{Li, et al.}

\begin{abstract}
Cross-media retrieval of scientific and technological information is one of the important tasks in the cross-media study. Cross-media scientific and technological information retrieval obtain target information from massive multi-source and heterogeneous scientific and technological resources, which helps to design applications that meet users’ needs, including scientific and technological information recommendation, personalized scientific and technological information retrieval, etc. The core of cross-media retrieval is to learn a common subspace, so that data from different media can be directly compared with each other after being mapped into this subspace. In subspace learning, existing methods often focus on modeling the discrimination of intra-media data and the invariance of inter-media data after mapping; however, they ignore the semantic consistency of inter-media data before and after mapping and media discrimination of intra-semantics data, which limit the result of cross-media retrieval. In light of this, we propose a scientific and technological information oriented Semantics-adversarial and Media-adversarial Cross-media Retrieval method (SMCR) to find an effective common subspace. Specifically, SMCR minimizes the loss of inter-media semantic consistency in addition to modeling intra-media semantic discrimination, to preserve semantic similarity before and after mapping. Furthermore, SMCR constructs a basic feature mapping network and a refined feature mapping network to jointly minimize the media discriminative loss within semantics, so as to enhance the feature mapping network's ability to confuse the media discriminant network. Experimental results on two datasets demonstrate that the proposed SMCR outperforms state-of-the-art methods in cross-media retrieval.
\end{abstract}

\keywords{cross-media retrieval, adversarial learning, scientific and technological information, media constraint }

\maketitle

\section{Introduction}
Science and Technology Information focuses on the cutting-edge trends of high-tech at home and abroad. Real-time follow-up of the latest scientific and technological information helps to promote the development of national strategic scientific and technological forces, drive scientific and technological innovation, and thus ensure high-quality national development. Scientific and technological information contains a large amount of multimedia information (such as images, texts, etc.), and has the characteristics of large volume, rich sources, and diverse types \cite{Peng2018Overview,Shi2021Collaborative,Li2017Recursive1}. As a current research hotspot, cross-media scientific and technological information retrieval is still faced with the problem that the heterogeneous gap and semantic gap between multimedia data need to be broken urgently \cite{Wei2016Dependent}. Through cross-media scientific and technological information retrieval, target scientific and technological information can be obtained from massive multi-source and heterogeneous scientific and technological resources \cite{LI2017Distributed}, so as to design applications that meet user needs, including scientific and technological information recommendation \cite{Li2018Implementation}, personalized scientific and technological information retrieval \cite{Salehi2015Examining,Yang2015Ontology}, etc.. We aims to solve the problem that the existing cross-media scientific and technological information retrieval methods only consider the discriminative loss of intra-media data and the invariance loss of inter-media data after mapping, while ignoring the semantic consistency loss of inter-media data before and after mapping, and the discriminative loss of intra-semantic data, which limit the effect of cross-media retrieval.

There are various types of cross-media scientific and technological information retrieval. Previous work \cite{Wang2013Coupled,Hu2020Anomaly,Hardoon2004Canonical,Wang2016Joint,Zhai2014Learning} focused on traditional statistical correlation analysis methods, learning a linear projection matrix in a common space by optimizing statistical values \cite{Gong2014embedding} to build a shared subspace, in which data belonging to different media can be directly compared with each other using common distances in this subspace. All of the above methods rely on a linear representation of the data; however, it is difficult to fully simulate the complex correlations of real-world cross-media data only by linear projection. Therefore, some studies \cite{Feng2014Retrieval,Yan2015matching,Peng2016Shared,KOU2016Social,Xu2016Image} solve the above problems through deep learning methods, using its powerful abstraction ability to deal with multi-layer nonlinear transformations of multimedia data for cross-media correlation learning. Existing deep learning-based cross-media retrieval models usually focus on preserving the pairwise similarity of coupled cross-media samples (e.g. images and texts) \cite{Ngiam2011Multimodal}. However, for one sample of one media, there may exist
more than one semantically different samples of the same media so
that this focus on pairwise coupled samples only is far from sufficient \cite{Wang2017Adversarial}. Recent works \cite{He2017Unsupervised,Li2018Adversarial,Wang2017Adversarial,Zhen2019Supervised,Liu2021Similarity} introduce the idea of adversarial learning to generate media-invariant representations for samples of different media in a common subspace by jointly performing label prediction and preserving the underlying cross-media semantic structure in the data. However, the above methods only focus on modeling the semantic discrimination of intra-media data and the semantic invariance of inter-media data after subspace mapping, while ignoring the semantic consistency of inter-media data before and after mapping, and the media discriminative within semantics. This limits the effect of cross-media retrieval.

To solve the above problems, we strengthens the ability to map different types of media data into a shared high-level semantic space by introducing inter-media semantic consistency and intra-semantics media constraints, and propose a scientific and technological information oriented Semantics-adversarial and Media-adversarial Cross-media Retrieval method (SMCR). SMCR adopts the idea of adversarial learning \cite{Goodfellow2014Generative} to construct a feature mapper and media discriminator for mini-max game. SMCR follows previous work \cite{Li2018Hashing,Yu2019Transfer} by utilizing label prediction to ensure that the data still retains intra-media distinctions after feature mapping. Different from previous work, SMCR simultaneously minimizes the distances between the data of different media in the same semantic text-image pair before and after feature mapping, respectively, to ensure the semantic consistency of the data between different media during the mapping process. To ensure the mapped data is semantically close to itself and far away from itself in media, we construct the basic mapping network and the refined mapping network to assist in modeling the intra-semantics media constraints. It helps to enhance the ability of the feature mapping network to confuse the media discrimination network. Meanwhile, the media discrimination network is responsible for distinguishing the original media of the mapped data, and once it is deceived, the entire game process converges.

We have three main contributions which are as follows:

\begin{enumerate}
	\item We propose a scientific and technological information oriented Semantics-adversarial and Media-adversarial Cross-media Retrieval (SMCR) to break the heterogeneous gap and semantic gap between multimedia data. SMCR effectively learns the public representation of heterogeneous data by maintaining the intra-media semantic discrimination, inter-media semantic consistency, and intra-semantics media discrimination in an end-to-end way;
	\item We model intra-semantics media constraints by constructing a basic feature mapping network and a refined feature mapping network to jointly perform feature mapping of multimedia data, to enhance the ability of the feature mapping network in confusing the media discrimination network;
	\item Extensive experiments on two datasets demonstrate that SMCR outperforms the current state-of-the-art cross-media retrieval methods, including traditional methods and deep learning-based methods.
\end{enumerate}

\section{related work}

Cross-media retrieval of scientific and technological information is a research hotspot in recent years, aiming to learn a common subspace \cite{Wang2016Joint,Wang2017Adversarial,XUE2019subspace,Xue2021Clustering}, in which data of different media can be directly compared with each other, so as to bridge the semantic gap between different media.

Classic methods of cross-media retrieval focus on statistical correlation analysis \cite{Wang2013Coupled,Hu2020Anomaly,Hardoon2004Canonical,Wang2016Joint,Zhai2014Learning}, which mainly learns the linear projection matrix of the common space by optimizing the statistical value. For instance, Hardoon et al. proposed Canonical Correlation Analysis (CCA) \cite{Hardoon2004Canonical} by correlating the linear relationship between two multidimensional variables, which can be viewed as using complex labels as a way to guide feature selection towards underlying semantics. This method exploits two perspectives of the same semantic object to extract semantic representations. Wang et al. proposed a Joint Feature Selection and Subspace Learning (JFSSL) \cite{Wang2016Joint}. Inspired by the underlying relationship between CCA and linear least squares, JFSSL utilizes coupled linear regression to learn the projection matrix such that the data for different media is mapped into a common subspace. At the same time, this method uses regularization to simultaneously select relevant and distinct features from different feature spaces, and uses multimedia graph regularization when mapping to preserve inter-media and intra-media similarity relationships. Zhai et al. \cite{Zhai2014Learning} proposed a novel feature learning algorithm for cross-media data called Joint Representation Learning (JRL). This method is able to jointly explore relevance and semantic information in a unified optimization framework, and integrate sparse and semi-supervised regularization for all media types into a unified optimization problem. It aims to simultaneously learn sparse projection matrices for different media and directly project raw heterogeneous features into the joint space. However, it is difficult to fully simulate the complex correlations of cross-media data in the real world through linear projection alone.

With the rise of deep learning, many studies have focused on applying deep neural networks capable of multi-layer nonlinear transformations to cross-media retrieval \cite{Feng2014Retrieval,Yan2015matching,Peng2016Shared,KOU2016Social,Xu2016Image}. For example, Yan et al. \cite{Yan2015matching} proposed a cross-media image caption matching method based on Deep Canonical Correlation Analysis (DCCA). By addressing non-trivial complexity and overfitting problems, this method is made suitable for high-dimensional image and text representations and large datasets. Peng et al. proposed a Cross-media Multiple Deep Network (CMDN) \cite{Peng2016Shared} to exploit complex and rich cross-media correlations through hierarchical learning. In the first stage, the CMDN does not only utilize separate representations within the media as previous works, however jointly learns two complementary separate representations for each media type; in the second stage, since each media type has two complementary separate representations, the method combines separate representations hierarchically in a deeper two-level network in order to jointly model inter-media and intra-media information to generate shared representations. However, existing deep neural network-based cross-media retrieval models usually only focus on preserving the pairwise similarity of coupled cross-media samples (e.g. an image and a piece of text), while ignoring one sample of one medium, which may exist multiple semantically different samples of the same media. Therefore, they cannot preserve cross-media semantic structure.

In recent years, state-of-the-art research on cross-media retrieval has turned to adversarial learning \cite{FANG2020Identity}. Although adversarial learning is widely used in image generation \cite{Radford2016UnsupervisedRL}, researchers also use it as a regularizer \cite{Ganin2015Domain}. Some studies have adopted the adversarial idea in cross-media retrieval and achieved remarkable results \cite{He2017Unsupervised,Li2018Adversarial,Wang2017Adversarial,Zhen2019Supervised,Liu2021Similarity}. For example, Wang et al. proposed Adversarial Cross-Modal Retriviel (ACMR) \cite{Wang2017Adversarial} to address the difficulty of preserving cross-media semantic structure. The method uses feature projectors to generate media-invariant representations for samples of different media in a common subspace by jointly performing label prediction and preserving the underlying cross-media semantic structure in the data. Its purpose is to confuse media classifier acting as an adversary. Media classifier tries to distinguish samples based on their media and in this way guide the learning of feature projectors. Through the convergence of this process, i.e. when the media classifier fails, the representation subspace is optimal for cross-media retrieval. Zhen et al. proposed a Deep Supervised Cross-modal Retrieval method (DSCMR) \cite{Zhen2019Supervised}, which aims to find a common representation space in which samples from different media can be directly compared. This method minimizes the discriminative loss in the label space and common representation space to supervise the model in learning discriminative features. At the same time, the media-invariant loss is minimized, and a weight-sharing strategy is used to remove cross-media differences of multimedia data in a common representation space to learn media-invariant features. Liu et al. proposed a Semantic Similarity based Adversarial Cross-media Retrieval method (SSACR) \cite{Liu2021Similarity}, using semantic distribution and similarity as the training basis of feature mapping network, so that the distance between different media data under the same semantics is small, and the distance between the same media data under different semantics is large. Finally, SSACR uses similarity to sort and obtain the search results in the same space. However, the above methods focus on modeling the semantic loss of intra-media data and the semantic loss of inter-media data after mapping, while ignoring the semantic consistency of inter-media data before and after mapping, and the media discrimination within semantics, which limit the effect of cross-media retrieval.

\begin{figure*}[ht]
	\centering
	\includegraphics[width=0.95\textwidth]{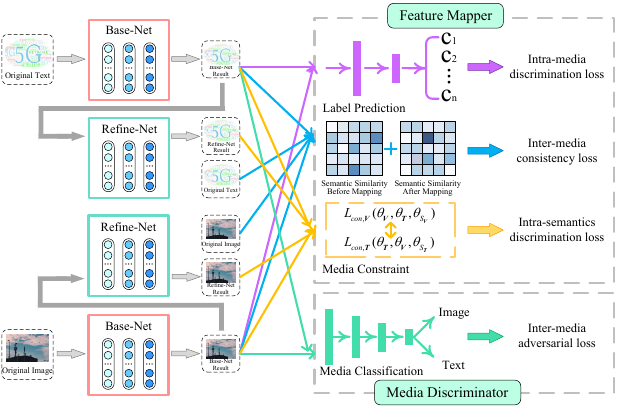}
	\caption{Graphical representation of our proposed SMCR. It consists of two processes that play the minimax game: a feature mapper generating intra-media semantics-discriminative, inter-media semantics-consistent, and intra-semantics media-discriminative representations, and a media discriminator distinguishing the original media of representations.}
	\label{fig:Graphical_SMCR}
\end{figure*}

\section{Problem Formulation}

There are various kinds of multimedia data. To keep generality, we focuses on the cross-media retrieval of text and images. Given a set of semantically related image-text pairs $m = \{ {m_1},{m_2}, \cdots, {m_{|m|}}\} $, where ${m_i} = ({v_i},{t_i})$ represents the $i$-th image-text pair in $m$, $v_i\in \mathbb{R}^{d_v}$ represents an image feature vector of dimension $d_v$, and $t_i\in \mathbb{R}^{d_t}$ represents a text feature vector of dimension $d_t$. Each image-text pair corresponds to a semantic category vector ${l_i} = [{y_1},{y_2}, \cdots {y_C}] \in \mathbb{R}^C $, where $C$ represents the total number of semantic categories. If ${l_i}$ belongs to the $j$-th semantic category, then record ${y_j} = 1$, otherwise record ${y_j} = 0$. We record the feature matrices corresponding to all images, texts, and semantic categories in $m$ as $V = \{ {v_1},{v_2}, \cdots, {v_N}\} \in \mathbb{R}^{{d_v} \times N} $, $T = \{ {t_1},{t_2}, \cdots, {t_N}\} \in \mathbb{R}^{{d_t} \times N} $, and $L = \{ {l_1},{l_2}, \cdots, {l_N}\} \in \mathbb{R}^{C \times N} $.

Our goal is to utilize data from one media (such as image $v_i$ or text $t_i$) to retrieve data from another media (such as text $t_i$ or image $v_i$). To compare the semantic similarity between different media data, we design two feature mapping network: Base-Net and Refine-Net. Base-Net map image features and text features into a unified latent semantic space $S$ for semantic similarity comparison. The image feature $V$ mapped to latent embedding space $S$ is denoted as ${S_V} = {f_V}(V;{\theta _V})$, and the text feature $T$ mapped to latent embedding space $S$ is denoted as ${S_T} = {f_T}(T;{\theta _T})$, where ${f_V}(V;{\theta _V})$ and ${f_T}(T;{\theta _T})$ represent the mapping functions of image and text, respectively. To further improve the quality of feature mapping, we use Refine-Net to map the output features of Base-Net. The image feature ${S_V}$ is mapped as $S^{'}_V = {g_{{S_V}}}({S_V};{\theta _{{S_V}}})$, and the text feature ${S_T} $ is mapped as $S^{'}_T = {g_{{S_T}}}({S_T};{\theta _{{S_T}}})$, where ${g_{{S_V}}}({S_V};{\theta _{{S_V}}})$ and ${g_{{S_T}}}({S_T};{\theta _{{S_T} }})$ represent the mapping function of image features and text features, respectively.

\section{Method}

We propose a scientific and technological information oriented Semantics-adversarial and Media-adversarial Cross-media Retrieval (SMCR). The framework of SMCR is shown in Fig.\ref{fig:Graphical_SMCR}. The motivation of SMCR is to use the idea of adversarial learning to continuously confront between semantics and media, and learn a common subspace, in which the data of different media can be directly compared with each other.

\subsection{Feature Mapping Network}

Feature mapping network is used to map the features of different media into a unified latent embedding space for the comparison of semantic similarity. At the same time, it also plays the role of the "generator" in GAN \cite{Goodfellow2014Generative}, which aims to confuse the media discriminant network (introduced in Section 4.2). In order to make the mapped feature representation fully consider the semantic similarity and media similarity of two types of media data, the feature mapping network consists of three parts: label prediction, semantics preservation, and media constraints. The label prediction within the media enables the features mapped in the latent embedding space to be semantically classified with the original semantic label as the true value; the semantics preservation between media enables the data of same semantics and different media to retain the semantic similarity; The media distinction within semantics makes the mapped data closer to the original semantics.

\subsubsection{Label Prediction}

To ensure the features mapped into the latent embedding space can still retain the original semantics, label prediction is performed with the original semantic labels as the ground truth. A softmax layer that maintains linear activations is added at the end of each feature map network. The image-text pairs are used as samples for training, and the output of the model is probability distribution of the semantic category corresponding to each data. We calculate intra-media discrimination loss by utilizing the loss function introduced in \cite{Wang2017Adversarial}:

\begin{equation}
{L_{imd}}({\theta _{imd}}) =  - \frac{1}{n}\sum\limits_{i = 1}^n {({l_i} \cdot (\log {{\widehat p}_i}({v_i}) + \log {{\widehat p}_i}({t_i})))} 
\end{equation}
where ${L_{imd}}$ represents the cross-entropy loss for semantic label prediction of all image-text pairs, ${\theta _{imd}}$ represents the parameters of the classifier, ${l_i}$ is the ground-truth of each sample ${m_i}$, and ${\widehat p_i}$ denote the result of probability distributions for each data (image or text) in the sample.

\subsubsection{Semantics Preservation}

The semantics preservation module is dedicated to ensuring that data with the same semantics and different media can retain the semantic similarity before and after mapping, i.e., the data with the same semantics in different media are closer, and the data with different semantics in different media are farther away. Before mapping to the latent embedding space $S$, the semantic distributions of image data and text data in each sample ${m_i}$ are ${l_v}$ and ${l_t}$, respectively. Then the loss of semantic consistency between two different media data is represented by the ${l_2}$ norm:

\begin{equation}
{l_2}(v,t) = {\left\| {{l_v} - {l_t}} \right\|_2}
\end{equation}

After mapping to the latent embedding space $S$, the semantic consistency loss between the image data feature ${S_V}$ and the text data feature ${S_T}$ in each sample ${m_i}$ are also calculated by ${l_2}$ norm:

\begin{equation}
{l_2}({\theta _V},{\theta _T}) = {\left\| {{f_V}(V;{\theta _V}) - {f_T}(T;{\theta _T})} \right\|_2}
\end{equation}

Therefore, the overall inter-media consistency loss can be modeled as the combination of ${l_2}(v,t)$ and ${l_2}({\theta _V},{\theta _T})$:

\begin{equation}
{L_{imi}}({\theta _V},{\theta _T}) = {l_2}(v,t) + {l_2}({\theta _V},{\theta _T})
\end{equation}
where ${L_{imi}}$ represents the loss of semantic consistency between media before and after mapping.

\subsubsection{Media Constraint}

In addition to facilitating the measurement of semantic similarity between different media data, another advantage of the feature mapping network is to generate mapped features to fool the media discrimination network, making it impossible to distinguish the original media of the feature. Therefore, a media constraint module within semantics is introduced. To map the features of indistinguishable media more realistically, in addition to the basic feature mapping network, i.e. Base-Net ${P_1}$, another feature mapping network ${P_2}$ with the same structure is constructed, which is called Refine-Net. The input of the Refine-Net ${P_2}$ is the output (${S_V}$ or ${S_T}$) of ${P_1}$, and the output of ${P_2}$ is $S^{'}_V=g_{S_V}(S_V;\theta_{S_V})$ or $S^{'}_T=g_{S_T}(S_T;\theta_{S_T})$, where $S^{'}_V$ and $S^{'}_T$ represent the mapped result of ${S_V}$ and ${S_T}$ which are mapped by Refine-Net ${P_2}$, ${g_{{S_V}}}( {S_V};{\theta _{{S_V}}})$ and $g_{S_T}({S_T};\theta_{S_T})$ represent mapping functions of two features ${S_V}$, ${S_T}$, respectively. 

For each image-text pair ${m_i}$, our goal is to make the feature ( $S^{'}_V$ or $S^{'}_T$) mapped by the Refine-Net ${P_2}$ away from the features (${S_V}$ or ${S_T}$) mapped by Base-Net ${P_1}$, while closing to the features (${S_T}$ or ${S_V}$) of same semantics. Inspired by \cite{Hoffer2015Metric,Liang2018Generative,Wei2018Learning}, intra-semantics discrimination loss is computed using the following constraint loss:

\begin{equation}
{L_{con,V}}({\theta_V},{\theta_T},{\theta_{{S_V}}}) = \max (0,{\left\| {S^{'}_V - {S_T}} \right\|_2} - {\left\| {S^{'}_V - {S_V}} \right\|_2})
\end{equation}

\begin{equation}
{L_{con,T}}({\theta_T},{\theta_V},{\theta_{{S_T}}}) = \max (0,{\left\| {S^{'}_T - {S_V}} \right\|_2} - {\left\| {S^{'}_T - {S_T}} \right\|_2})
\end{equation}

Therefore, the overall intra-semantics media constraint loss can be modeled as combination of the constraint loss  ${L_{con,V}}({\theta _V},{\theta _T},{\theta _{{S_V}} })$ of image media data and the constraint loss ${L_{con,T}}({\theta _T},{\theta _V},{\theta _{{S_T}}})$ of text media data:

\begin{equation}
{L_{con}}({\theta _V},{\theta _T},{\theta _{{S_V}}},{\theta _{{S_T}}}) = {L_{con,V}}({\theta _V},{\theta _T},{\theta _{{S_V}}}) + {L_{con,T}}({\theta _T},{\theta _V},{\theta _{{S_T}}})
\end{equation}

\subsubsection{Feature Mapper}

In summary, the mapping loss of the entire feature mapping network is composed of intra-media discrimination loss, inter-media consistency loss, and intra-semantics discrimination loss, denoted as ${L_{emb}}$:

\begin{equation}
{L_{emb}}({\theta _V},{\theta _T},{\theta _{{S_V}}},{\theta _{{S_T}}},{\theta _{imd}}) = \alpha  \cdot {L_{imi}} + \beta  \cdot {L_{con}} + {L_{imd}}
\end{equation}
where $\alpha$ and $\beta$ are adjustable parameters to control the participation of the two types of losses in the entire feature mapping network's loss.

\subsection{Media Discriminator}

The media discriminant network plays the role of the “discriminator” in GAN \cite{Goodfellow2014Generative} to determine the original media of the data mapped to the latent embedding space. Let the data label passing through the image mapping function be 0, and the data label passing through the text mapping function be 1. Wer use a three-layer fully connected network with a parameter of ${\theta _D}$ as the discriminant network, acting as the adversary of the feature mapping network. Its goal is to minimize the media classification loss, also known as the adversarial loss ${L_{adv}}$, defined as:

\begin{equation}
{L_{adv}}({\theta _D}) =  - \frac{1}{n}\sum\limits_{i = 1}^n {(\log D({v_i};{\theta _D}) + \log (1 - D({t_i};{\theta _D})))} 
\end{equation}
where ${L_{adv}}$ represents the cross entropy loss of each sample ${m_i}$ in the media discriminant network, $D( \cdot ;{\theta _D})$ represents the resulting media probability distribution of each data (image or text).

\subsection{Adversarial Learning}

The purpose of adversarial learning is to learn the optimal feature representation network parameters by simultaneously minimizing the mapping loss of Eq. (8) and the adversarial loss of Eq. (9). Adversarial learning consists of two sub-processes:

\begin{equation}
\begin{aligned}
&({\theta _V},{\theta _T},{\theta _{{S_V}}},{\theta _{{S_T}}},{\theta _{imd}})  \\ 
&=\mathop {\arg \min }\limits_{{\theta _V},{\theta _T},{\theta _{{S_V}}},{\theta _{{S_T}}},{\theta _{imd}}} ({L_{emb}}({\theta _V},{\theta _T},{\theta _{{S_V}}},{\theta _{{S_T}}},{\theta _{imd}}) - {L_{adv}}({\theta _D}))
\end{aligned}
\end{equation}

\begin{equation}
{\theta _D} = \mathop {\arg \max }\limits_{{\theta _D}} ({L_{emb}}({\theta _V},{\theta _T},{\theta _{{S_V}}},{\theta _{{S_T}}},{\theta _{imd}}) - {L_{adv}}({\theta _D}))
\end{equation}

The specific adversarial learning training process of SMCR is shown in Algorithm \ref{alg:Adversarial_learning}.

\begin{algorithm}[htb]
	\caption{Adversarial training process of SMCR}
	\label{alg:Adversarial_learning}
	\KwIn{Image feature set $V= \{ {v_1},{v_2}, \cdots,{v_N} \}$, text feature set  $T=\{ {t_1},{t_2}, \cdots,{t_N}\}$, ground-truth semantic label set $L = \{ {l_1},{l_2}, \cdots ,{l_N}\} $, iteration number $k$, learning rate $\mu$, data volume $m$ for each mini-batch, loss parameter $\lambda$.}
	\KwOut{Parameters $\theta_V$, $\theta_T$, $\theta_{S_V}$, $\theta_{S_T}$.}
	\BlankLine
	Initialize the parameters of model randomly. \\
	\While{not converge}{
		\For{$iter$=1 ot $k$}{
			update parameters $\theta_V$, $\theta_T$, $\theta_{S_V}$, $\theta_{S_T}$, ${\theta _{imd}}$ by descending their stochastic gradients \\
			${\theta _V} \leftarrow {\theta _V} - \mu  \cdot {\nabla _{{\theta _V}}}{\textstyle{1 \over m}}({L_{emb}} - {L_{adv}})$ \\
			${\theta _T} \leftarrow {\theta _T} - \mu  \cdot {\nabla _{{\theta _T}}}{\textstyle{1 \over m}}({L_{emb}} - {L_{adv}})$ \\
			${\theta _{{S_V}}} \leftarrow {\theta _{{S_V}}} - \mu  \cdot {\nabla _{{\theta _{{S_V}}}}}{\textstyle{1 \over m}}({L_{emb}} - {L_{adv}})$ \\
			${\theta _{{S_T}}} \leftarrow {\theta _{{S_T}}} - \mu  \cdot {\nabla _{{\theta _{{S_T}}}}}{\textstyle{1 \over m}}({L_{emb}} - {L_{adv}})$ \\
			${\theta _{imd}} \leftarrow {\theta _{imd}} - \mu  \cdot {\nabla _{{\theta _{imd}}}}{\textstyle{1 \over m}}({L_{emb}} - {L_{adv}})$ \\
		}
	}
	update parameters by ascending its stochastic gradients. \\
	${\theta _D} \leftarrow {\theta _D} + \mu  \cdot \lambda  \cdot {\nabla _{{\theta _D}}}{\textstyle{1 \over m}}({L_{emb}} - {L_{adv}})$
\end{algorithm}

\section{Experimental Setup}

\subsection{Research Questions}

\begin{itemize}
	\item[\textbf{RQ1}:]
	Does our Semantics-adversarial and Media-adversarial Cross-media Retrieval method SMCR outperform the state-of-the-art baseline algorithms?
	\item[\textbf{RQ2}:]
	What is the contribution of the key ingredient of SMCR for cross-media retrieval?
	\item[\textbf{RQ3}:]
	Is the performance of SMCR sensitive to parameters?
\end{itemize}

\subsection{Dataset}

To answer the above research questions, experiments were conducted using a dataset crawled from the science and technology information website SciTechDaily\footnote{https://scitechdaily.com/news/technology}. The dataset consists of 5,217 image-text pairs, 4,173 of which are used as training set and 1,044 as test set. To verify the generality of the SMCR model, experiments are also conducted using the Wikipedia \cite{Costa2014Correlation} dataset. The Wikipedia dataset consists of 2866 image-text pairs, 2,292 of which are used as the training set and 574 as the test set. The details of the two datasets are shown in Table \ref{tab:dataset}.

\begin{table}[htb]
	\centering
	\caption{Properties of the two datasets used for the experiments}
	\begin{tabular}{ccccc}
		\hline
		dataset          & sample        & label & image feature      & text feature      \\
		\hline
		SciTechDaily & 4,173/1,044 & 8  & 4,096d VGG & 6,500d BoW \\
		\hline
		Wikipedia    & 2,292/574  & 10 & 4,096d VGG & 5,000d BoW \\
		\hline
	\end{tabular}
	\label{tab:dataset}
\end{table}

\subsection{Baselines}

We compares SMCR with the following baselines and state-of-the-art algorithms:

\noindent
\hangafter 1
\hangindent 2em
$\textbf{\textit{Canonical Correlation Analysis (CCA)}}$: This model \cite{Hardoon2004Canonical} learns a common subspace for data of different media types to maximize the pairwise correlation between two sets of heterogeneous data.

\noindent
\hangafter 1
\hangindent 2em
$\textbf{\textit{Joint Feature Selection and Subspace Learning (JFSSL)}}$: This model \cite{Wang2016Joint} learns a projection matrix to map multimedia data into a common subspace, and simultaneously selects related features and distinguishing features from different feature spaces.

\noindent
\hangafter 1
\hangindent 2em
$\textbf{\textit{Cross-media Multiple Deep Network (CMDN)}}$: This model \cite{Peng2016Shared} exploits complex cross-media correlations through hierarchical learning. In the first stage, intra-media and media information are jointly modeled; in the second stage, inter-media representations and intra-media representations are hierarchically combined to further learn rich cross-media correlations.

\noindent
\hangafter 1
\hangindent 2em
$\textbf{\textit{Adversarial Cross-Modal Retriviel (ACMR)}}$: This model \cite{Wang2017Adversarial} seeks efficient common subspaces based on adversarial learning. A triple constraint is imposed on the feature projector to minimize the gap between the representations of all samples from different media with the same semantic label, while maximizing the distance between semantically different images and texts.

\noindent
\hangafter 1
\hangindent 2em
$\textbf{\textit{Deep Supervised Cross-modal Retrieval (DSCMR)}}$: This model \cite{Zhen2019Supervised} is also based on the idea of adversarial learning. The discriminative loss in the label space and the common representation space is minimized, while the media invariance loss is minimized, and a weight sharing strategy is used to eliminate cross-media differences of multimedia data in the common representation space.

\noindent
\hangafter 1
\hangindent 2em
$\textbf{\textit{Semantic Similarity based Adversarial Cross-media Retrieval (SSACR)}}$: This model \cite{Liu2021Similarity} is also based on the idea of adversarial learning. The similarity of the feature vectors of different media data mapped to the same semantic space is calculated and compared with the similarity between the original semantic feature vectors to eliminate the differences between different media data under the same semantics.

\noindent
To further analyze the contribution of the key ingredients of our SMCR to cross-media retrieval, we use two versions of SMCR for baselines:

\noindent
\hangafter 1
\hangindent 2em
$\textbf{\textit{SMCR(without L}}$$_{\textbf{\textit{imi}}}$$\textbf{\textit{)}}$: This model is the SMCR model which removes the inter-media semantic loss $L_{imi}$.

\noindent
\hangafter 1
\hangindent 2em
$\textbf{\textit{SMCR(without L}}$$_{\textbf{\textit{con}}}$$\textbf{\textit{)}}$: This model is the SMCR model which removes the intra-semantics media loss $L_{con}$.

\begin{table*}[ht]
	\centering
	\caption{Comparison of cross-media retrieval performance on SciTechDaily and Wikipedia datasets}
	\begin{tabular}{|c|c|ccc|ccc|ccc|} 
		\hline
		\multicolumn{2}{|l|}{~}                             & \multicolumn{3}{c|}{mAP@5}                          & \multicolumn{3}{c|}{mAP@25}                         & \multicolumn{3}{c|}{mAP@50}                          \\ 
		\hline
		\multicolumn{1}{|c|}{dataset}           & method        & txt2img         & img2txt         & Average         & txt2img         & img2txt         & Average         & txt2img         & img2txt         & Average          \\ 
		\hline
		\multirow{7}{*}{\rotatebox{90}{SciTechDaily}} & CCA           & 0.2337          & 0.1806          & 0.2071          & 0.2328          & 0.1761          & 0.2044          & 0.2225          & 0.1789          & 0.2007           \\
		& JFSSL         & 0.3984          & 0.2852          & 0.3418          & 0.3817          & 0.2777          & 0.3297          & 0.3699          & 0.2647          & 0.3173           \\
		& CMDN          & 0.4483          & 0.3514          & 0.3998          & 0.4299          & 0.3443          & 0.3871          & 0.4206          & 0.3229          & 0.3717           \\
		& ACMR          & 0.5131          & 0.4382          & 0.4756          & 0.4943          & 0.4471          & 0.4707          & 0.4966          & 0.4259          & 0.4612           \\
		& DSCMR         & 0.5042          & 0.4577          & 0.4809          & 0.4812          & 0.4646          & 0.4729          & 0.4810          & \textbf{0.4467} & 0.4638           \\
		& SSACR         & 0.5091          & 0.4572          & 0.4831          & 0.5049          & 0.4487          & 0.4768          & 0.5072          & 0.4355          & 0.4713           \\
		& \textbf{SMCR} & \textbf{0.5270} & \textbf{0.4790} & \textbf{0.5030} & \textbf{0.5291} & \textbf{0.4727} & \textbf{0.5009} & \textbf{0.5191} & 0.4426          & \textbf{0.4808}  \\ 
		\hline
		\multirow{7}{*}{\rotatebox{90}{Wikipedia}}    & CCA           & 0.2639          & 0.2154          & 0.2396          & 0.2883          & 0.2255          & 0.2569          & 0.2575          & 0.2152          & 0.2363           \\
		& JFSSL         & 0.4432          & 0.3481          & 0.3956          & 0.4266          & 0.3528          & 0.3897          & 0.4152          & 0.3479          & 0.3815           \\
		& CMDN          & 0.5265          & 0.4194          & 0.4729          & 0.5046          & 0.4171          & 0.4608          & 0.4874          & 0.3938          & 0.4406           \\
		& ACMR          & 0.6372          & 0.4920          & 0.5646          & 0.6251          & 0.4937          & 0.5594          & 0.5887          & 0.4824          & 0.5355           \\
		& DSCMR         & 0.6413          & 0.4963          & 0.5688          & 0.6514          & 0.5082          & 0.5798          & 0.6452          & \textbf{0.4973} & 0.5712           \\
		& SSACR         & 0.6642          & 0.4927          & 0.5784          & 0.6608          & \textbf{0.5089} & 0.5848          & 0.6416          & 0.4956          & 0.5686           \\
		& \textbf{SMCR} & \textbf{0.7014} & \textbf{0.5059} & \textbf{0.6036} & \textbf{0.6714} & 0.5003          & \textbf{0.5858} & \textbf{0.6503} & 0.4959          & \textbf{0.5731}  \\
		\hline
	\end{tabular}
	\label{tab:Comparison}
\end{table*}

\subsection{Evaluation Metrics}

We use the classic evaluation metric in cross-media retrieval \cite{Kou2019Common,Liang2020Semantic,Liang2019Grained}, mean Average Precision (mAP), to evaluate the performance of algorithms in the two tasks: using text to retrieve image (txt2img) and using image to retrieve text (img2txt). To calculate mAP, we need to calculate the average precision of $R$ retrieved documents $AP = {\textstyle{1 \over T}}\sum\nolimits_{r = 1}^R {P(r)\delta (r )} $ first, where $T$ is the number of related documents in the retrieved documents, $P(r)$ represents the precision of the first $r$ retrieved documents, if the $r$th retrieved document is relevant, then $ \delta (r) = 1$, otherwise $\delta (r) = 0$. Then mAP is calculated by averaging the AP values of all queries in the query set. The larger the mAP value, the more accurate the results of cross-media retrieval.

\section{Results and Analysis}

In this section, all experimental results are analyzed to answer the research questions raised in Section 5.1.

\subsection{Effectiveness of SMCR}

To answer \textbf{\textit{RQ1}}, we compare SMCR with six state-of-the-art algorithms on two datasets, SciTechDaily and Wikipedia, respectively. The baselines are: 1) methods based on statistical correlation analysis: CCA\cite{Hardoon2004Canonical}, JFSSL\cite{Wang2016Joint}; 2) methods based on deep learning: CMDN\cite{Peng2016Shared}, ACMR\cite{Wang2017Adversarial}, DSCMR\cite{Zhen2019Supervised}, SSACR\cite{Liu2021Similarity}.

Table \ref{tab:Comparison} shows the mAP values (mAP@5, mAP@25, mAP@50) calculated for the first 5, 25, and 50 retrieval results on the two tasks of using text to retrieve image (txt2img) and using image to retrieve text (img2txt). Besides, it also demonstrate the mean results of mAP (Average) on the two retrieval tasks. From Table 2, we have the following findings:

1) SMCR outperforms all state-of-the-art algorithms, including methods based on statistical correlation analysis and methods based on deep learning. It is worth mentioning that the mean mAP of the SMCR algorithm on the first 5, 25, and 50 retrieval results is superior to the current state-of-the-art SSACR algorithm on both datasets. This demonstrates that although SSACR also models intra-media semantic loss and inter-media semantic loss, SMCR introduces inter-media semantic consistency loss and intra-semantics discrimination loss, which help to further improve cross-media retrieval performance by more realistically mapping the feature representation of indistinguishable media;

2) SMCR, JFSSL, CMDN, ACMR, DSCMR, SSACR, which simultaneously model intra-media similarity and inter-media similarity, are better than CCA modeling inter-media similarity based on image-text pair. This finding indicates that considering both intra-media similarity and inter-media similarity can improve the performance of cross-media retrieval;

3) The cross-media retrieval performance of SMCR, ACMR, DSCMR, and SSACR is better than that of CMDN, which also models inter-media invariance and intra-media discrimination in the multi-task learning framework, indicating that adversarial learning helps to further improve the modeling effect of inter-media invariance and intra-media discrimination;

4) SMCR, which separately modeling the semantic similarity of different media data with the same semantics before and after mapping, outperforms ACMR and DSCMR, which only model the semantic similarity of different media data with the same semantics after mapping. This finding demonstrates that modeling the semantic invariance of data from different media before and after mapping helps to improve the performance of cross-media retrieval;

5) There are consistent performance of SMCR and all state-of-the-art algorithms on both SciTechDaily and Wikipedia datasets, which indicates that the SMCR algorithm is not only limited to the retrieval of cross-media scientific and technological information, but also has good performance in general cross-media retrieval tasks.

\begin{figure*}[ht] 
	\centering
	\includegraphics[width=0.95\textwidth]{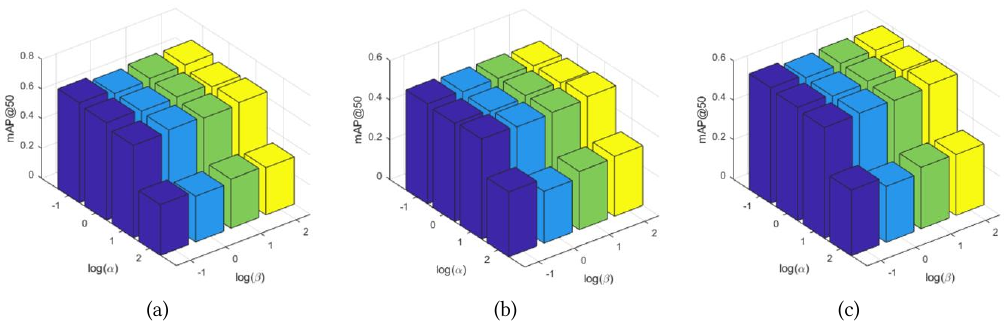}
	\caption{The performance of (a) using text to retrieve image, (b) using image to retrieve text, and (c) the average retrieval performance with various parameter and in Wikipedia dataset.}
	\label{fig:parameter}
\end{figure*}

\subsection{Contribution of Key Ingredients of SMCR}

Next, to answer \textbf{\textit{RQ2}}, we compare SMCR with two versions of SMCR: SMCR without inter-media consistency loss ${L_{imi}}$, and SMCR without intra-semantics discrimination loss ${L_{con}}$. Since the intra-media discrimination loss ${L_{imd}}$ modeled by label prediction is not an innovation in our SMCR, the SMCR without ${L_{imd}}$ is not compared.

\begin{table}
	\centering
	\caption{Performance of SMCR and its variants in SciTechDaily.}
	\begin{tabular}{|c|c|ccc|} 
		\hline
		method   & mAP    & txt2img & img2txt & Average  \\ 
		\hline
		\multirow{3}{*}{\begin{tabular}[c]{@{}c@{}}SMCR \\(without $L_{imi}$)\end{tabular}} & mAP@5  & 0.5196  & 0.4627  & 0.4911    \\
		& mAP@25 & 0.5187  & 0.4525  & 0.4856  \\
		& mAP@50 & 0.5024  & 0.4408  & 0.4716  \\ 
		\hline
		\multirow{3}{*}{\begin{tabular}[c]{@{}c@{}}SMCR \\(without $L_{con}$)\end{tabular}} & mAP@5  & 0.5155  & 0.4513  & 0.4834    \\
		& mAP@25 & 0.5073  & 0.4474  & 0.4773  \\
		& mAP@50 & 0.4972  & 0.4386  & 0.4679  \\ 
		\hline
		\multirow{3}{*}{SMCR}                                                                        & mAP@5  & 0.5270  & 0.4790  & 0.5030  \\
		& mAP@25 & 0.5291  & 0.4727  & 0.5009    \\
		& mAP@50 & 0.5191  & 0.4426  & 0.4808    \\
		\hline
	\end{tabular}
	\label{tab:Variants_SciTechDaily}
\end{table}

Tables \ref{tab:Variants_SciTechDaily} and \ref{tab:Variants_Wikipedia} show the comparison results of SMCR and its two variants on the SciTechDaily and Wikipedia datasets, respectively. We have the following findings:
1) SMCR without inter-media consistency loss ${L_{imi}}$ and SMCR without intra-semantics discrimination loss ${L_{con}}$, underperform SMCR in cross-media retrieval. This finding demonstrates that optimizing the inter-media consistency loss ${L_{imi}}$ and the intra-semantics discrimination loss ${L_{con}}$ simultaneously in the feature mapping network is more helpful to improve the cross-media retrieval performance than optimizing one of them alone;
2) There are consistent cross-media retrieval performance of SMCR and its variants on both SciTechDaily and Wikipedia datasets. Again, this finding indicates that the SMCR algorithm is not only limited to cross-media scientific and technological information retrieval, but also effective on general cross-media retrieval tasks.

\begin{table}
	\centering
	\caption{Performance of SMCR and its variants in Wikipedia.}
	\begin{tabular}{|c|c|ccc|} 
		\hline
		method    & mAP    & txt2img & img2txt & Average  \\ 
		\hline
		\multirow{3}{*}{\begin{tabular}[c]{@{}c@{}}SMCR \\(without $L_{imi}$)\end{tabular}} & mAP@5  & 0.6919  & 0.4983  & 0.5951   \\
		& mAP@25 & 0.6622  & 0.4937  & 0.5779   \\
		& mAP@50 & 0.6418  & 0.4901  & 0.5659   \\ 
		\hline
		\multirow{3}{*}{\begin{tabular}[c]{@{}c@{}}SMCR \\(without $L_{con}$)\end{tabular}} & mAP@5  & 0.6806  & 0.5038  & 0.5922   \\
		& mAP@25 & 0.6596  & 0.4980  & 0.5788   \\
		& mAP@50 & 0.6416  & 0.4938  & 0.5677   \\ 
		\hline
		\multirow{3}{*}{SMCR}                                                                   & mAP@5  & 0.7014  & 0.5059  & 0.6036   \\
		& mAP@25 & 0.6714  & 0.5003  & 0.5858   \\
		& mAP@50 & 0.6503  & 0.4959  & 0.5731   \\
		\hline
	\end{tabular}
	\label{tab:Variants_Wikipedia}
\end{table}

\subsection{Parameter Sensitivity of SMCR}

Finally, we answer \textbf{\textit{RQ3}}. The mapping loss ${L_{emb}}$ of the feature mapping network in Eq. (8) has two parameters $\alpha $ and $\beta $, which control the participation of inter-media consistency loss ${L_{imi}}$ and intra-semantics discrimination loss ${L_{con}}$ in the overall mapping loss ${L_{emb}}$. This section changes the values of $\alpha $ and $\beta $ on the Wikipedia dataset to test the parameter sensitivity of the SMCR algorithm. The parameters $\alpha $ and $\beta $ are set to 0.1, 1, 10, 100 respectively. In particular, when $\alpha = 0$, SMCR degenerates into SMCR without inter-media consistency loss ${L_{imi}}$; when $\beta = 0$, the SMCR degenerates into an SMCR without intra-semantics discrimination loss ${L_{imi}}$. Therefore, the values of $\alpha $ and $\beta $ are not 0. Under the premise of fixing one parameter (such as $\alpha $), we change another parameter (such as $\beta $) to conduct experiments, and use mAP@50 value to evaluate the performance of using text to retrieve image, using image to retrieve text, and average retrieval effect respectively. The results are shown in Fig. \ref{fig:parameter}.

As can be seen from Fig. \ref{fig:parameter}, when the value of $\alpha $ is {0.1, 1, 10}, and the value of $\beta $ is {0.1, 1, 10, 100}, SMCR performs better. This indicates that SMCR is insensitive to parameters, i.e. has better generalization ability. In particular, on the task of using text to retrieve image, SMCR performs best when $\alpha = 0.1$ and $\beta = 0.1$; on the task of using image to retrieve text, when $\alpha = 1$ and $\beta = - 1$, SMCR achieves the best retrieval effect; in terms of average retrieval effect, when $\alpha = - 1$ and $\beta = - 1$, SMCR performs best.

\section{Conclusion}

We proposes a Scientific and Technological Information Oriented Semantics-adversarial and Media-adversarial Cross-media Retrieval method (SMCR), which can simultaneously learn intra-media discriminative, inter-media consistent, and intra-semantics discriminative representations in cross-media retrieval. SMCR is based on an adversarial learning approach and involves two processes in the minimax game: feature mapping networks generate representations with intra-media discrimination, inter-media consistency, and inter-semantics discrimination, and media discrimination networks try to discriminate the original media of given data. SMCR introduces inter-media consistency loss to ensure that the data between media before and after the mapping retains semantic consistency; in addition, the intra-semantic media discriminative loss is introduced to ensure that the mapped data is semantically close to itself, and far away from itself in media, to enhance The ability of feature mapping networks to confuse media discrimination networks. Experimental results on two cross-media datasets demonstrate the effectiveness of the SMCR algorithm and the performance of SMCR outperform state-of-the-art methods on cross-media retrieval.

\begin{acks}
This work was supported by National Key R\&D Program of China (2018YFB1402600), the National Natural Science Foundation of China (61772083, 61802028). To my mother, Hong Li, for supporting my work and saving most of the time for me to finish my work. To all the enemies who can't kill me for inspiring my courage.
\end{acks}

\bibliographystyle{ACM-Reference-Format}
\bibliography{mybibliography}

\end{document}